\documentclass{article}
\usepackage{amsmath}
\usepackage{amsfonts}
\usepackage{amssymb}

\begin{document}

\title{Contribution of One-Time Pair Correlation Function\\to Kinetic Phenomena in Nonequilibrium Gas}
\author{A. Ya. Shul'man\\Institute of Radioengineering and Electronics of the RAS\\Moscow 101999, Russia. E-mail: ash@cplire.ru}
\date{}
\maketitle

\begin{abstract}
It has been established in nineteen seventies that in nonequilibrium case the
pair collisions generate non-zero two-particle correlations which are
non-diagonal in momentum space and give the essential contribution to the
current fluctuations of hot electrons. It is shown here that this correlations
give also a contribution to the collision integral, i.e., to kinetic
properties of nonequilibrium gas. The expression for electron energy loss rate
$P$ via phonons is re-derived in detail from this point of view. The
contribution of the non-diagonal part of the nonequilibrium pair correlator to
phonon-electron collision integral and to $P$ is obtained and explicitly
calculated in the electron temperature approximation. It is shown that these
results can be obtained from stochastic non-linear kinetic equation with
Langevin fluctuation force. Such an approach allows to formulate the simple
general conditions under that a contribution of two-particle correlations
might be essential in kinetics. The contribution obtained does not contain the
extra powers of small gas parameter unlike the equilibrium virial decompositions.
\end{abstract}

\footnotetext{Based on a poster presented at the conference ``Progress in
Nonequilibrium Green's Functions, Dresden, Germany, 19.-22. August 2002''}

\section{Introduction}

The derivation of kinetic equation for one-particle distribution function
$F(p,r,t)$ from BBGKY chain leads to the collision integral expressed in terms
of the two-particle correlation function. The latter is in turn expressed by
$F(p,r,t)$ using Boltzmann's Sto\ss zahlansatz that implies the neglect of
initial one-time correlation function $g_{2}\left(  p_{1},p_{2},r,t\right)  $
, which is non-diagonal in the momentum space. In nonequilibrium case,
however, it has been established \cite{GGK69}$-$\cite{ASH70} that pair
collisions generate the non-diagonal two-particle correlator that gives the
essential contribution to the current fluctuations of hot electrons. Here we
show that this correlator gives also a contribution to the collision integral,
i.e., to kinetic properties of nonequilibrium electrons. The expression for
electron energy loss rate $P$ via phonons is re-derived in detail from this
point of view.

In fact, the formula for $P$ in terms of electron density-density correlation
function $\left\langle \rho_{k}\rho_{k}^{+}\right\rangle _{\omega}$ was
obtained by Kogan \cite{ShK62} using Fermi Golden Rule. However, to come to
the well-known result for $P$ in the electron temperature approximation the
fluctuation-dissipation theorem was implicated to express $\left\langle
\rho_{k}\rho_{k}^{+}\right\rangle _{\omega}$ in terms of the dielectric
function of the electron gas. This step is defensible in the case of the true
equilibrium electron gas only and it gives rise to results equivalent to the
Sto\ss zahlansatz. In this report the expression for $P$ in terms of
$\left\langle \rho_{k}\rho_{k}^{+}\right\rangle _{\omega}$ is derived from
quantum kinetic equation for phonons interacting with nonequilibrium electron
gas. The contribution of the non-diagonal part of the nonequilibrium pair
correlator to phonon-electron collision integral and to $P$ is obtained and
calculated explicitly in the electron temperature approximation.

It is shown that these results may be obtained from stochastic non-linear
kinetic equation with Langevin fluctuation force. Such an approach allows to
formulate the simple general conditions when the two-particle correlator
contribution might be essential in kinetics. The derivation of this equation
based on the procedure of partial averaging over microscopic-scale
fluctuations similar to the procedure using by H. Mori \cite{Mori73} to derive
the generalized classical Boltzmann equation for dense gases and liquids.
However, in our case the contribution obtained does not contain the extra
powers of small gas parameter of the kinetic theory.

\section{Kinetic equation for phonons}

\subsection{Basic designations}

Hamiltonian $H=H_{e}+H_{ph}+V_{eph}$

Electron part $H_{e}=\sum\limits_{k}\hbar\Omega_{k}a_{k}^{+}a_{k}+V_{ee}$
where $V_{ee}$ stands for the electron-electron interaction energy

Phonon part $H_{ph}=\sum\limits_{f}\hbar\omega_{f}b_{f}^{+}b_{f}$

Electron-phonon interaction

$V_{eph}=\frac{1}{\sqrt{v_{0}}}\sum\limits_{f}\left(  c_{f}\rho_{f}^{+}%
b_{f}+c_{f}^{\ast}\rho_{f}b_{f}^{+}\right)  $

Spacial Fourier components of the electron density operator

$\rho_{f} =\sum\limits_{k}a_{k+\frac{f}{2} }^{+} a_{k-\frac{f}{2} }
,\,\,\,\,\,\,\,\,\,\,\,\rho_{f}^{+} =\rho_{-f} $

Phonon distribution function

$N_{f} (t)=\left\langle \left(  b_{f}^{+} b_{f} \right)  _{t} \right\rangle $

Electron distribution function

$F_{k} (t)=\left\langle \left(  a_{k}^{+} a_{k} \right)  _{t} \right\rangle $

The angle brackets denote the full averaging with the density matrix
corresponding to the noninteracting electrons and phonons at an initial
instant of time. The subscript $t$ indicates the Heisenberg representation for
electron and phonon operators.

\subsection{Phonon-electron collisional integral\newline and the electron
density-density correlation function}

We obtain the kinetic equation for slow variables $N_{f}(t)$, smoothing
Heisenberg equations of motion over microscopically-large time interval
$\Delta t$. Such a technique was suggested by N.N. Bogolyubov and N.M. Krylov
(see the book \cite{Bog2}, pp. 5-76) and it has been used in \cite{Thesis70}
for the derivation of the linearized kinetic equation for the occupation
number fluctuations with Langevin sources of fluctuations. With accuracy to
the second-order terms in $e-ph$ interaction we have
\begin{equation}
\frac{\Delta N_{f}(t)}{\Delta t}=-\frac{1}{\hbar^{2}\Delta t}\left\langle
S^{+}(t)\int\limits_{t}^{t+\Delta t}dt^{\prime}\int\limits_{t}^{t\prime
}dt^{\prime\prime}\left[  V_{eph}^{0}(t^{\prime\prime}),\left[  V_{eph}%
^{0}(t^{\prime}),(b_{f}^{+}b_{f})_{t}\right]  \right]  S(t)\right\rangle
\label{KEgeneral}%
\end{equation}
The superscript $0$ denotes operators in the interaction representation with
respect to the electron-phonon interaction $V_{ep}$.

Let us remind interrelations between $S$-matrix and the time evolution of any
operator $\hat{O}$ in the Heisenberg and interaction representations:
\begin{align}
U(t,t_{0})  &  \equiv e^{-\frac{i}{\hbar}H(t-t_{0})}=U^{0}(t,t_{0}%
)S(t,t_{0}),\,U^{0}(t,t_{0})=e^{-\frac{i}{\hbar}(H_{e}+H_{ph})(t-t_{0}%
)},\nonumber\\
\hat{O}(t)  &  =U^{+}(t,t_{0})\hat{O}(t_{0})U(t,t_{0})=S^{+}(t,t_{0})\hat
{O}^{0}(t,t_{0})S(t,t_{0})\label{Sdefinit}\\
i\hbar\partial S/\partial t  &  =V_{eph}^{0}(t)S,\,S(t_{0})=1,\,\hat{O}%
^{0}(t,t_{0})=U^{0+}(t,t_{0})\hat{O}(t_{0})U^{0}(t,t_{0})\nonumber
\end{align}

Calculations of the internal commutators in Eq.(\ref{KEgeneral}) lead to the
desired result involving the electron density-density correlation function:
\begin{equation}
\frac{dN_{f}}{dt}=\frac{\left|  c_{f}\right|  ^{2}}{v_{0}\hbar^{2}}\left\{
\left\langle \rho_{f}^{+}\rho_{f}\right\rangle _{-\omega_{f}}^{0}\left(
N_{f}+1\right)  -\left\langle \rho_{f}\rho_{f}^{+}\right\rangle _{\omega_{f}%
}^{0}N_{f}\right\}  \label{KEwRo}%
\end{equation}
where the commutativity of free electron and phonon operators in
Eq.(\ref{KEgeneral}) and the condition $\omega_{f}\Delta t\gg1$ have been used.

The function
\begin{equation}
\left\langle \rho_{f}\rho_{f}^{+}\right\rangle _{\omega}=\int\limits_{-\infty
}^{\infty}d\tau\left\langle \rho_{f}(t+\tau)\rho_{f}^{+}(t)\right\rangle
\exp\left(  i\omega\tau\right)  \label{DDcorr}%
\end{equation}
is the spectral density of the correlation function of the electron density
fluctuations. It is interested to remark that the right hand side of
Eq.(\ref{KEwRo}) may be rather interpreted as the emission and the absorption
of phonons by the electron fluid with the structure factor $\left\langle
\rho_{f}\rho_{f}^{+}\right\rangle _{\omega_{f}}^{0}$ than as individual
electron-phonon collisions.

\subsection{Spectral density of the electron fluctuations}

The sought spectral density must be calculated without $V_{ep}$ taken into
account. From equations of motion for the electron operators we obtain in the
random phase approximation:
\begin{equation}
\left\langle \rho_{f}^{+}\rho_{f}\right\rangle _{\omega}^{0}=\frac
{\left\langle \rho_{f}^{+}\rho_{f}\right\rangle _{\omega}^{00}}{\left|
\epsilon\left(  \omega,f\right)  \right|  ^{2}}, \label{DDobserv}%
\end{equation}
where $\epsilon(\omega,f)$ is the dielectric function of the electron gas and

\begin{multline}
\left\langle \rho_{f}\rho_{f}^{+}\right\rangle _{\omega}^{00}=\sum
\limits_{k,k_{1}}\left\langle \rho_{f}(k)\rho_{f}^{+}(k_{1})\right\rangle
_{t}\nonumber\\
\times\left[  \frac{1}{i\left(  \omega_{k+\frac{f}{2}}-\omega_{k-\frac{f}{2}%
}-\omega\right)  +\eta}+\frac{1}{-i\left(  \omega_{k_{1}+\frac{f}{2}}%
-\omega_{k_{1}-\frac{f}{2}}-\omega\right)  +\eta}\right]  \label{DDindepend}%
\end{multline}
is the spectral density of the density-density correlation function for
non-interacting electrons. The one-time correlation function is
\begin{equation}
\left\langle \rho_{f}(k)\rho_{f}^{+}(k_{1})\right\rangle _{t}\equiv
\left\langle a_{k-\frac{f}{2}}^{+}a_{k+\frac{f}{2}}a_{k_{1}+\frac{f}{2}}%
^{+}a_{k_{1}-\frac{f}{2}}\right\rangle _{t}.
\end{equation}
The one-time correlation function of the non-interacting electrons is
independent of time (in steady state) under condition that the electron
energies $\hbar\Omega_{k}=\hbar^{2}k^{2}/2m$ obey the equality
\begin{equation}
\Omega_{k-\frac{f}{2}}-\Omega_{k+\frac{f}{2}}+\Omega_{k_{1}+\frac{f}{2}%
}-\Omega_{k_{1}-\frac{f}{2}}\equiv\frac{\hbar}{m}f\left(  k_{1}-k\right)  =0.
\end{equation}

As long as $f\neq0$ it is necessary to be $k_{1}=k$ . Therefore, this
condition provides the one-time correlator to be the integral of motion during
the time interval $\Delta t$ like the distribution function $F_{k}$.

\subsection{Kinetic equation with account for the contribution of
nonequilibrium pair correlations}

The substitution of the explicit expression for correlators to the kinetic
equation (\ref{KEwRo}) gives rise to%

\begin{equation}%
\begin{tabular}
[c]{l}%
$\frac{dN_{f}}{dt}=$\\
$\frac{2\left|  C_{f}\right|  ^{2}}{v_{0}\hbar^{2}\left|  \epsilon\left(
\omega_{f},f\right)  \right|  ^{2}}\sum\limits_{k}\left\langle a_{k+\frac
{f}{2}}^{+}a_{k-\frac{f}{2}}a_{k-\frac{f}{2}}^{+}a_{k+\frac{f}{2}%
}\right\rangle _{t}\delta\left(  \Omega_{k+\frac{f}{2}}-\Omega_{k-\frac{f}{2}%
}-\omega_{f}\right)  \left(  N_{f}+1\right)  -$\\
$-\frac{2\left|  C_{f}\right|  ^{2}}{v_{0}\hbar^{2}\left|  \epsilon\left(
\omega_{f},f\right)  \right|  ^{2}}\sum\limits_{k}\left\langle a_{k-\frac
{f}{2}}^{+}a_{k+\frac{f}{2}}a_{k+\frac{f}{2}}^{+}a_{k-\frac{f}{2}%
}\right\rangle _{t}\delta\left(  \Omega_{k+\frac{f}{2}}-\Omega_{k-\frac{f}{2}%
}-\omega_{f}\right)  N_{f}.$%
\end{tabular}
\label{KEw<>}%
\end{equation}

Taking into account the commutation rules $a_{k_{1}}^{+}a_{k_{1}}+a_{k_{1}%
}a_{k_{1}}^{+}=1$ it is easy to obtain
\begin{equation}
\left\langle a_{k}^{+}a_{k_{1}}a_{k_{1}}^{+}a_{k}\right\rangle _{t}%
=F_{k}(t)-\left\langle a_{k}^{+}a_{k_{1}}^{+}a_{k_{1}}a_{k}\right\rangle _{t}.
\end{equation}
According to the results from Gantsevich et al. \cite{GGK69}
\begin{equation}
\left\langle a_{k}^{+}a_{k_{1}}^{+}a_{k_{1}}a_{k}\right\rangle =F_{k}F_{k_{1}%
}+\phi\left(  k,k_{1}\right)  \label{one-time}%
\end{equation}
where $\phi\left(  k,k_{1}\right)  $ is a nonequilibrium correction to the
pair correlator.

Finally, the kinetic equation for phonons with corrections for the
non-equilibrium part of the electron pair correlator takes the form:%

\begin{equation}%
\begin{tabular}
[b]{l}%
$%
\begin{array}
[c]{c}%
\frac{dN_{f}}{dt}=\\
\frac{2\left|  C_{f}\right|  ^{2}\left(  N_{f}+1\right)  }{v_{0}\hbar
^{2}\left|  \epsilon\left(  \omega_{f},f\right)  \right|  ^{2}}\sum
\limits_{k}\left[  F_{k}\left(  1-F_{k-f}\right)  -\phi\left(  k,k-f\right)
\right]  \delta\left(  \omega_{k}-\omega_{k-f}-\omega\right)  -\\
-\frac{2\left|  C_{f}\right|  ^{2}N_{f}}{v_{0}\hbar^{2}\left|  \epsilon\left(
\omega_{f},f\right)  \right|  ^{2}}\sum\limits_{k}\left[  F_{k}\left(
1-F_{k+f}\right)  -\phi\left(  k,k+f\right)  \right]  \delta\left(  \omega
_{k}-\omega_{k+f}+\omega\right)  .
\end{array}
$%
\end{tabular}
\label{KEwPhi}%
\end{equation}

It should be stressed that for ordinary classical equilibrium electron gas
there are no pair correlations between different states in momentum space,
i.e., $\phi\left(  k,k_{1}\right)  \equiv0$. This fact can be easily seen from
the form of the Gibbs distribution with the usual kinetic energy part of the
Hamilton's function $K=\sum_{i}\mathbf{p}_{i}^{2}/2m$ since there are no terms
proportional to $\mathbf{p}_{i}\mathbf{p}_{j}$ ($i$,$j$ stand for the particle enumeration).

\section{Interrelation between non-linear stochastic kinetic equation and
correlation corrections}

Following \cite{Mori73}$-$\cite{ASH79} and using partial averaging over fast
and short-wave fluctuations under conditions that the slow variables like
$N_{f}(t)$ and $F_{k}(t)$ are constant instead of the full averaging over all
states of the system under consideration, we find the non-linear stochastic
kinetic equation with Langevin sources of fluctuations in the form
\begin{equation}%
\begin{tabular}
[c]{l}%
$\frac{d\tilde{N}_{f}}{dt}=J_{f}(t)+\frac{2\left|  C_{f}\right|  ^{2}}%
{v_{0}\hbar^{2}\left|  \epsilon\left(  \omega_{f},f\right)  \right|  ^{2}%
}\times$\\
$\{\left(  \tilde{N}_{f}+1\right)  \sum\limits_{k}\left[  \tilde{F}_{k}\left(
1-\tilde{F}_{k-f}\right)  \right]  \delta\left(  \omega_{k}-\omega
_{k-f}-\omega\right)  -$\\
$-\tilde{N}_{f}\sum\limits_{k}\left[  \tilde{F}_{k}\left(  1-\tilde{F}%
_{k+f}\right)  \right]  \delta\left(  \omega_{k}-\omega_{k+f}+\omega\right)
\}$%
\end{tabular}
\label{KEstoch}%
\end{equation}
The similar stochastic kinetic equation can be derived for the fluctuating
electron occupation numbers $\tilde{F}_{k}$ with accounting for $V_{ee}$ and
$V_{eph}$ interactions \cite{ASH79}. Both equations are nonlinear in
$\tilde{F}_{k}$ and bilinear in $\tilde{N}_{f}\tilde{F}_{k}$. Here tilde
denotes that the slow fluctuations are still kept. Langevin source of
fluctuations can be expressed as the difference between the exact time
derivative of the phonon occupation number and its approximation

$J_{f}(t)=\frac{i}{\hbar\Delta t}S^{+}(t)\int\limits_{t}^{t+\Delta
t}dt^{\prime}\left[  V_{eph}^{0}(t^{\prime}),N_{f}^{0}(t^{\prime})\right]
S(t)-\frac{i}{\hbar\Delta t}S^{+}(t)\int\limits_{t}^{t+\Delta t}dt^{\prime
}\widetilde{\left[  V_{eph}^{0}(t^{\prime}),N_{f}^{0}(t^{\prime})\right]  }S(t)$

It is seen that the partial average of Langevin sources themselves is equal to
zero $\tilde{J}_{f}(t)=0$ . The partial correlation function $\frac{1}%
{2}\widetilde{[J_{f}(t_{1}),J_{q}(t_{2})]_{+}}$ is expressed by means of the
transitions' rates involved in the collision integral similarly to the case of
the linear stochastic kinetic equation for fluctuations of the occupation
numbers \cite{KSH69},\cite{Thesis70}.

Taking full average from stochastic kinetic equation and using the evident
properties (see Eq.(\ref{one-time}))
\begin{equation}
\left\langle \tilde{F}_{k}\right\rangle =F_{k};\,\,\,\left\langle \tilde
{F}_{k}\tilde{F}_{k_{1}}\right\rangle =F_{k}F_{k_{1}}+\phi(k,k_{1}),k\neq
k_{1},\text{and also }\left\langle J_{f}(t)\right\rangle =0
\end{equation}
we can easy obtain the same kinetic equation for phonon occupation numbers
with correlation corrections that was derived above. Moreover, from this point
of view we can immediately write the respective terms in the Boltzmann kinetic
equation for neutral particles and bring up the question about a possible
contribution of nonequilibrium particle-particle correlations to the equations
describing phenomena like flow turbulence. The similar statement concerns to
electron-phonon correlations described by expressions like $\left\langle
\delta\tilde{F}_{k}\delta\tilde{N}_{f}\right\rangle _{t}$ in the case when a
phonon nonequilibrium takes place.

In conclusion of these general considerations it needs to underline that the
contribution of non-equilibrium correlations to kinetics should be expected in
that cases when the excitations considered are scattered by nonequilibrium
many-particle system.

\section{Electron energy loss rate via phonons}

The energy loss rate of electrons due to phonons per unit volume can be
obtained using the relation:
\begin{equation}
P=\frac{1}{v_{0}}\sum\limits_{f}\hbar\omega_{f}\frac{dN_{f}}{dt}.
\end{equation}
It is seen from the kinetic equation for phonons Eq.(\ref{KEwPhi}) that the
loss rate can be broken up into two parts:

\begin{equation}
P=P_{1}+P_{2}, \label{Ptotal}%
\end{equation}
where on the right hand side the first term is related to the one-particle
electron distribution function and it is described by the known expressions.
The second one is due to the correlation contributions determined by the
function $\phi$. We assume during the calculations of $P$ that the phonons are
in thermal equilibrium and the electron-electron collisions are sufficiently
frequent in order to provide the use of the electron temperature approximation
for hot electrons. As a result of these assumptions the following formulae may
be derived:
\begin{align}
P_{1}(T)  &  =\sum\limits_{f}\frac{2\left|  C_{f}\right|  ^{2}}{\left(
v_{0}\hbar\right)  ^{2}\left|  \varepsilon\left(  \omega_{f},f\right)
\right|  ^{2}}\hbar\omega_{f}\left[  N_{f}^{0}(T)-N_{f}\right] \nonumber\\
&  \times\sum\limits_{k}\left(  F_{k}^{0}-F_{k+f}^{0}\right)  \delta\left(
\Omega_{k+f}-\Omega_{k}-\omega_{f}\right)  \label{P1}%
\end{align}
and
\begin{equation}
P_{2}(T,T_{0})=-\sum\limits_{f}\frac{2\left|  C_{f}\right|  ^{2}}{(v_{0}%
\hbar)^{2}\left|  \epsilon(\omega_{f},f\right|  ^{2}}\hbar\omega_{f}%
\sum\limits_{k}\phi\left(  k+f,k\right)  \delta\left(  \Omega_{k+f}-\Omega
_{k}-\omega_{f}\right)  . \label{P2}%
\end{equation}

Here $N_{f}^{0}(T)=\left[  \exp\left(  \hbar\omega_{f}/T\right)  -1\right]
^{-1}$ is the equilibrium Bose-Einstein distribution with hot-electron
temperature $T$ and $N_{f}$ is the true nonequilibrium phonon distribution
function. Further, we will consider the equilibrium phonons, i.e.,
$N_{f}=N_{f}^{0}\left(  T_{0}\right)  $. In this case the expression
(\ref{P1}) for $P_{1}$ coincides with well-known Kogan's formula except for
dielectric function in the denominator describing the dynamic screening of the
electron-phonon interaction.

The function $\phi\left(  k,k_{1}\right)  $ was found in \cite{ASH70} under
accepted assumptions and can be presented in the form:
\begin{equation}
\phi(k,k_{1})=Q(T,T_{0})\frac{\partial F_{k}^{0}}{\partial T}\frac{\partial
F_{k_{1}}^{0}}{\partial T},\label{PhiTe}%
\end{equation}
where $F_{k}^{0}(T)$ is the Maxwell distribution of non-degenerate hot
electrons and
\begin{equation}
Q(T,T_{0})=\frac{T^{2}}{nc_{e}}\left[  \frac{P(T,T_{0})/\left(  T-T_{0}%
\right)  }{dP/dT}-1\right]  \label{Qmultip}%
\end{equation}

The expression for $\phi$ in \cite{ASH70} was obtained without accounting for
the contribution of $\phi$ itself in $P$ . Therefore, the quantity $P_{1}$ was
only implied as the total energy loss rate $P$. Now evidently we have the
equation for determination of $P_{2}(T,T_{0})$ since this quantity is defined
by the right hand side of the equality (\ref{P2}) where $\phi(k+f,k)$ depends
itself on $P_{2}(T,T_{0})$ accordingly to Eqs.(\ref{Ptotal}),(\ref{PhiTe}%
-\ref{Qmultip}). The situation is very resembling the self-consistent field
approach in the plasma theory. Now we have to solve simultaneously the set of
kinetic equations for the one-particle distribution function and the pair correlator.

It needs to stress that the substitution of the total loss rate $P$ instead of
$P_{1}$ in the expression (\ref{PhiTe}) for $\phi\left(  k,k_{1}\right)  $ is
justified because all symmetry properties of the electron-electron and
electron-phonon integrals of collisions used in \cite{ASH70} during the
derivation are not destroyed by introducing $\phi\left(  k,k_{1}\right)  $ in
collision integrals for electrons and phonons.

It is of importance to note in conclusion that the expression (\ref{PhiTe})
for $\phi\left(  k,k_{1}\right)  $ found in the limit of high frequency of the
electron-electron collisions does contain coupling constants of neither
electron-electron nor electron-phonon interactions. Therefore, the
corresponding corrections to the integral of phonon-electron collisions do not
contain additional powers of the small parameters of the kinetic theory. At
low heating it is proportional to $T-T_{0}$ like $P_{1}(T,T_{0})$, as it can
be obtained from the expression (\ref{Qmultip}) for $Q(T,T_{0})$. However, the
solution (\ref{PhiTe}) for $\phi\left(  k,k_{1}\right)  $ has been obtained in
\cite{ASH70} in the case of the non-degenerate electron gas. Hence $F_{k}%
^{0}(T)\ll1$ and the product of two distribution functions in Eq.(\ref{PhiTe})
makes $P_{2}\ll P_{1}$ in the case of scattering electrons by acoustic phonons
since $\hbar\omega_{f}\ll T,T_{0}$ and $N_{f}\gg1$ for phonons involved. The
situation can change in the case of low-temperature scattering of electrons by
optical phonons under condition $\hbar\omega_{0}\gg T,T_{0}$. The $\phi\left(
k,k_{1}\right)  $ corrections should be more pronounced for the hot-electron
phenomena in the degenerate electron gas of semiconductors or metals
\cite{HEM94}.

\section*{Acknowledgments}

The partial financial support of Russian Foundation for Basic Researches and
Deutsche Forschungsgemeinschaft is gratefully acknowledged.

\end{document}